# PhantomOS: A Next Generation Grid Operating System

Irfan Habib[2], Kamran Soomro[2], **Ashiq Anjum**[1], Richard McClatchey[1], Arshad Ali[2], Peter Bloodsworth[1]

1  CCS Research Centre, University of the West of England, Bristol, UK,
{ashiq.anjum, richard.mcclatchey,peter.bloodsworth}@cern.ch
2  National University of Sciences and Technology, Rawalpindi, Pakistan
{irfan.habib,arshad.ali, kamran.soomro}@niit.edu.pk

## Abstract

Grid Computing has made substantial advances in the past decade; these are primarily due to the adoption of standardized Grid middleware. However Grid computing has not yet become pervasive because of some barriers that we believe have been caused by the adoption of middleware centric approaches. These barriers include: scant support for major types of applications such as interactive applications; lack of flexible, autonomic and scalable Grid architectures; lack of plug-and-play Grid computing and, most importantly, no straightforward way to setup and administer Grids. PhantomOS is a project which aims to address many of these barriers. Its goal is the creation of a user friendly pervasive Grid computing platform that facilitates the rapid deployment and easy maintenance of Grids whilst providing support for major types of applications on Grids of almost any topology. In this paper we present the detailed system architecture and an overview of its implementation.

## 1. Introduction

Grid computing has made massive progress during the last decade. However, some barriers to adoption remain which have been primarily created by the middleware approach to Grid computing. Such difficulties have significantly restricted the pervasiveness of Grid technologies [1]. Currently Grids are far from easy to deploy or maintain and we feel that improvements must be made in order to improve the take up of Grid technologies.

Deployment of Grids with existing Grid middleware [2,3] involves the installation of multiple layers of software. Most of these do not support all types of computations; they have scant support for interactive applications popular in common user centric environments, such as in so-called Health Grids [4]. Development of common Grid applications is complex and most existing development environments provide the application developer with a Grid that is non-transparent [5]. Existing Grid middleware can create inflexible network topologies. Often these topologies are not fault tolerant as required in highly dynamic service oriented Grids. The Globus Toolkit, one of the leading Grid middleware, does not address environments characterized by very large and highly dynamic Grid membership [6].

The focus of current Grid Middleware is to act as the glue between clusters of participating organizations. Substantial amounts of manual and static configuration are required on the part of the system administrators. As a consequence Grid middleware has a steep learning curve. The difficulty in setting up Grid middleware also restricts the uptake of Grid computing in non-research environments. Mathews et al [7] have highlighted similar issues. Multiple software layers in a Grid do not ensure fault tolerance. For example, with the popular cluster execution service Condor [8], a centralized cluster middleware can be liable to complete failure if a central server crashes. Active research is being pursued into more robust, flexible and fault tolerant Grid architectures, by converging Grid and Peer to Peer (P2P) topologies. However no Grid as of yet, has shown the advantages of such convergence.

In the light of all the above stated limitations, it is clear that in order to facilitate the adoption of Grid computing to new domains and make it user-friendly for existing users' latent drawbacks in its architecture must be addressed. PhantomOS, our proposed Grid Operating system, aims at developing a pervasive grid computing platform which addresses the drawbacks of the existing infrastructure, leading to a fault tolerant, flexible and easy to use stack for rapid deployment of Grids.

PhantomOS aims to transparently grid-enable all types of modern computations from traditional batch oriented to interactive, without requiring customization to applications in order to deploy them over a Grid. It is an operating system with built in support for Grid Computing. It is also an integrated Grid stack, that builds on and extends existing Grid technologies to enable rapid deployment of Grids, and enabling "plug and play" Grid computing on a fault tolerant resource discovery architecture. This paper proceeds as follows: firstly we outline some related efforts in Section 2. In Section 3 and 4 we describe the architecture and implementation status of individual components of the architecture. In Section 5 we conclude and highlight future work.

## 2. Related Work

Recently two major efforts in the direction of Grid operating systems (Grid OS) have been launched: Vigne and XtreemOS. The Vigne Grid Operating System [9] is a Grid OS which aims to relieve users and programmers from the burden of dealing with the highly distributed and volatile resources of computational grids. Vigne focuses on three issues: i) Grid level single system images to provide abstractions for users and programmers to hide physical distribution of grid resources; ii) self-healing services to tolerate failure and reconfigurations in the Grid and iii) self-organization to relieve administrators from manually configuring and maintaining Vigne OS's services. Vigne plugs onto the Kerrighed Cluster system [10] which supports cluster middleware level issues. However Kerrighed has some limitations which would limit wide scale deployment. Kerrighed does not tolerate node failure; clusters cannot be bigger than 32 nodes and provide no symmetric multiprocessing and 64 bit architecture support.

XtreemOS [11] aims at the design and implementation of an open source Grid operating system with native support for virtual organizations (VO) which would be capable of running on a wide range of underlying platforms, from clusters to mobiles. XtreemOS plans to implement a set of system services to extend those found in a typical Linux system. These services will provide Grid computing capabilities to individual nodes. The aims of XtreemOS are similar to the PhantomOS project, both are Linux based and open source and both try to develop an OS level Grid solution with support for grid enabling applications and providing self-healing services for large scale dynamic Grids. PhantomOS focuses on an end-to-end stack for Grid computing for PCs and clusters, whereas XtreemOS focuses additionally on small-scale mobile devices as well as supporting applications ranging from eScience, finance, and eCommerce to multimedia.

Apple xGrid [12] is a part of the Apple Mac OSX operating system, which enables an organization to create a Compute Grid or compute cluster. Apple xGrid is perhaps one of the first common-user oriented Grid computing systems. Jobs submitted by a user to an Apple xGrid system are divided into independent tasks by the 'Controller', a machine set up to coordinate the computations on the Grid. The tasks are dispatched to 'Agents' which are dedicated machines offering their computing resources for the execution of tasks. Apple xGrid has some drawbacks in that the division of the Grid into Agents, Controllers, and submission machines is unscalable due to the client server nature of its interaction. Furthermore, xGrid has not been deployed in environments with large numbers of machines in multiple domains which would give a true indication of its scalability. Apple xGrid is not self-organizing, which might be the single most important hurdle to its transition towards a universal Grid platform.

## 3. Architecture of PhantomOS

PhantomOS aims at enabling the deployment of Grids which support any type of computations running on any topology. Figure 1 displays the main architectural components. PhantomOS is based on a two tier super-peer based paradigm [13]. Super peer paradigms have recently gained popularity as they enable Grids to integrate some of the advantages of peer to peer systems making a Grid infrastructure more robust, scalable and fault tolerant [14].

Architectural components of PhantomOS are modular in nature. Components are designed to be self-configuring and plug-and-play in order to facilitate the rapid deployment of a Grid e.g. adding a node to a site involves a simple sign-in procedure, adding a site to a region, involves a simple registration process with a region peer.

The OS can be seen from two perspectives: Firstly an integrated Grid Stack allowing rapid deployment of Grids, whilst making administration of Grids less complex. It can also be seen as an operating system which provides built in support for Grid computing. Figure 1 shows PhantomOS from both perspectives, the components which have a dotted background are those which are relevant to PhantomOS as a Grid Stack; others are for PhantomOS as a complete Operating System. There is overlap between both modes. Kernel changes can be turned off by unloading the appropriate kernel modules. If an organization chooses to use the stack configuration they can easily unload the kernel space modifications and use Grid computing from a user and middleware level at the cost of loosing support for Grid enabled interactive applications.

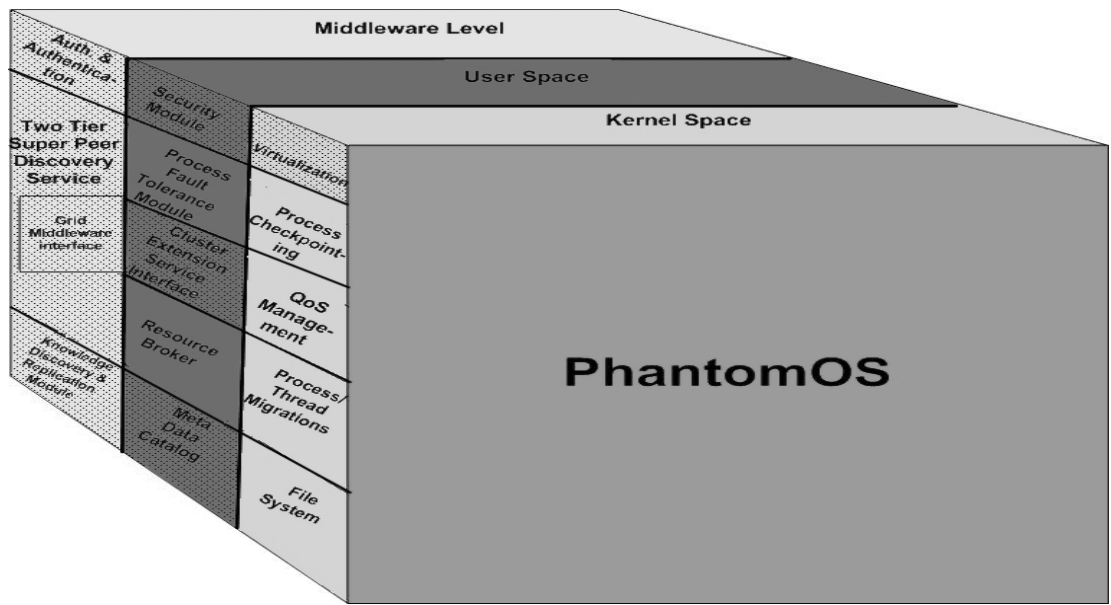

Figure 1: PhantomOS Architecture

The lowest layer in PhantomOS (shown in Figure 1 as the first vertical layer) is the kernel layer and includes modules which facilitate grid enabling of interactive application and fine-grained resource management in Grids. Few technologies support transparent grid enabling of interactive applications over a Grid. PhantomOS however makes use of process migration which transfers the execution context of processes to nodes where enhanced processing capabilities are available. Process migrations allow the transparent grid-enabling of existing applications without any need to modify them. More fine-grained thread migration and shared memory over a site or high-speed backbone is also supported.

Support for virtualization [15, 16] is another central feature of the kernel layer. PhantomOS aims to investigate hardware based virtualization in order to use a virtualization hypervisor which is both lightweight and efficient to enable the rapid creation and destruction of on-demand virtual machines. Both the QoS Management and kernel level process checkpointing modules are specifically designed for interactive applications. A Quality of service (QoS) Management module allows users to regulate resource usage of applications and to autonomously migrate them to different nodes within a site. The Checkpointing module allows execution states to be saved, so when failures occur the processes can be resumed.

The User Space layer contains components which allow PhantomOS to run existing Grid applications. It will also contain services which extend existing cluster middleware like Condor, to be self-healing, self-configuring and fault tolerant. Other crucial components in this layer include the Resource broker for interactive applications as well as the fault tolerance module which wraps the kernel checkpointing functionality and makes it available to the user applications. The Security module builds on the kernel level virtualization hypervisor and allows users to configure its behaviour. The Middleware level includes components which allow PhantomOS nodes to self organize into sub-grids, analogous to clusters and to provide resource discovery for both Grid applications and fine grained interactive applications. Interoperability with the existing Grid infrastructure through standardized interfaces both in terms of resource discovery and authentication and authorization will also be provided in the near future.

## 4. Implementation Status

### 4.1 Grid Enabling Interactive Applications

One of the core ideas behind PhantomOS is to support most types of applications including interactive applications, which are not easily supported in the current Grid middleware. Transparently grid-enabling interactive applications could remove some of the development related obstacles, such as the need to customize applications for the Grid leading to an easier way of developing Grid applications. However, as has been the custom in previous distributed OS projects [18], we do not want to provide an API for migration and process control. This would lead a user to customize applications specifically for the OS. Instead of this we want to support a transparent mechanism where developers do not have to specifically develop for the Grid.

The PhantomOS mechanism of Grid enabling interactive applications takes advantage of the fact that interactive applications are multi-threaded in nature. The kernel of Grid OS will support thread migration, an extension of the concept of process migration [19], which is popular in cluster middleware. The Grid OS kernel will be capable of migrating a single thread to another system, which will be selected by the broker as the best available site for execution. Migration will be initiated if an executing application exceeds the resource limits defined by the user, in a quality of service specification. One welcome consequence of supporting kernel level migration is that no changes in the user level are required; hence legacy applications can be run on the Grid without requiring modifications.

PhantomOS also provides for checkpointing of remotely migrated processes in order to save execution states and restart them in the case of an event. In figure 1. this is the job of the Process Fault Tolerance Module. The frequency of checkpointing, since it is pure overhead, is proportional to the instability of the machine. On a machine that has a history of many failures, the checkpoints will be made more frequently than on a machine with a more stable history. The created checkpoints will also be exported to the parent node of the process at regular intervals. However, the frequency of exports will be less than the frequency of checkpointing on the local machine, in order to contain the network cost incurred when transferring process checkpoints. The following formula has been used to calculate future checkpoint intervals and is itself a function of previous intervals.

$$I_n = W*I_{n-1} + (1-W)I_{n-2,}$$

where the value of W is dictated by site level policy.

Each new checkpointing interval is a function of previous checkpointing intervals along with a constant, W, which determines the importance a subgrid needs to give to the most recent checkpoint interval.

Currently PhantomOS builds on process migrations provided by OpenMOSIX [20], an open source cluster operating system. However there are numerous issues with OpenMOSIX's process migration mechanism: There is no support for thread migration or shared memory. Also it is built on Linux 2.4.26 kernel, which is relatively old and does not provide support for the latest hardware such as SATA drives, and kernel level virtualization (see section 4.3). Therefore porting and extending the migration mechanism to Linux 2.6.20 and above is 'work in progress'. Future PhantomOS nodes will standardize on Linux 2.6.20 and will provide thread migration support as well as shared memory. As the number of processor cores increases, multithreaded application development will become important. To support them over a Grid, shared memory is thus a prerequisite. Shared memory also simplifies the creation of grid applications as existing frameworks already support multithreading.

### 4.2 Fine Grained Resource Management and Security

Existing Grid middleware delegates resource management to the Cluster middleware. Modern Cluster middleware provides "all or nothing" resource control: a node is either completely available for processing, or it is not. To allow for fine grained control over users' resources, PhantomOS provides two technologies. PhantomOS makes use of OS virtualization to control resource usage and to provide security from foreign computation, by creating virtual machines which use the amount of resources the user is willing to provide to external users and makes use of a QoS management module for local computations. The QoS Management Module is used to force the migration of processes which exceed the user defined QoS limits.

The QoS module is implemented as a kernel module which monitors real-time resource usage of processes, with support for multi-core processors, and a daemon service to track trends of resource usage over time. This allows the system to make intelligent decisions about processes which are exceeding administrator defined resource limitations. Currently the QoS mechanism provides support for controlling CPU and memory utilization; furthermore networking usage will be provided in future releases. The QoS can be turned off, if the PhantomOS node is to be used as part of a dedicated Grid.

In PhantomOS we plan to use KVM, a kernel level virtualization hypervisor which has been recently added to the Linux kernel. KVM has a distinct edge over other virtualization engines. It takes advantage of built in hardware virtualization offered in the latest processors. Future processors are likely to have built-in virtualization support. Machines which are not running virtualization enabled hardware will not be able to take advantage of the PhantomOS security mechanism, and will have to revert to traditional methods of security. Figure 2 demonstrates the main concepts.

The host OS, O-Local, runs the virtual machine monitor which multiplexes the system resources between multiple competing guest OS instances. O-Remote, the guest OS, is a virtual machine for remote users. There is complete separation between them to ensure privacy for the remote users, and to ensure security for the local ones. The O-Local instance has complete access to the system. The O-Remote, however will have access only as required to complete the execution of a job, and access to I/O devices will be forbidden for this instance.

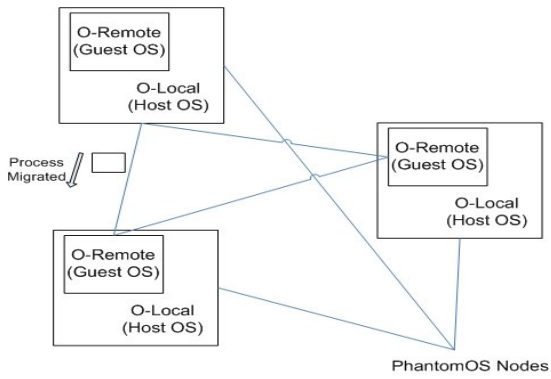

Figure 2: Virtualization in PhantomOS

### 4.3 Resource Broker

Our resource brokering algorithm is a decentralized peer-to-peer (P2P) network compute and data aware algorithm (NCDA) and considers both network connectivity and computational capability in scheduling decisions. It can be applied both at the local site and large-scale levels. It is a variant of the simple Condor ClassAds[20] based algorithm. ClassAds and related resource brokering and scheduling policies (henceforth abbreviated as FLOPS based algorithms) take the computational capability and the existing load into consideration and are popular in Cluster middleware. When a job is scheduled for migration to the Grid the resource broker polls members of the local sub grid for their current resource states. The member nodes respond to the request by dynamically creating resource descriptions and additionally they try at runtime to determine the current real bandwidth from the machine which requested the descriptions.

NCDA was benchmarked against the FLOPS based algorithms, and the round robin algorithm, which is popular in homogenous distributed computing environments and is implemented in some Grid middleware, most notably the Chimera VDS [21]. We define the performance of an algorithm as the quality of the resource selection and execution; if an algorithm has "high performance" it has the capability to select the best possible resources for a job.

We have used SimGrid 3.1[22] to implement the algorithms, and test them in randomly generated computing platforms of 50 to 1000 nodes with random network links ranging from 56kps to 10MBps; the computational capability of the nodes generated ranges from 10KFLOPS to 100MFLOPS. Applications were generated for each of the three categories: compute intensive, network intensive and hybrid applications, in each instance 1000 jobs were scheduled. Compute intensive applications require the processing of one Giga floating point operations each, whereas network intensive applications caused network traffic of 1GByte each. Hybrid applications required both processing of 1 GFLOPS and caused 1 GB network traffic each. The graphs in figures 4, 5 and 6 show the performance of each algorithm in each instance of scheduling.

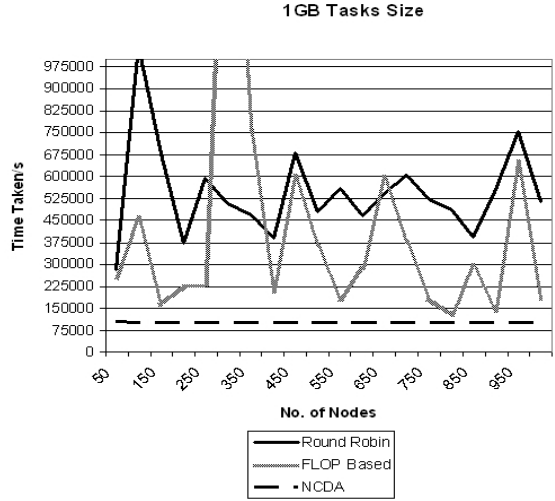

Figure 3: Performance of Algorithms for Network Intensive Applications

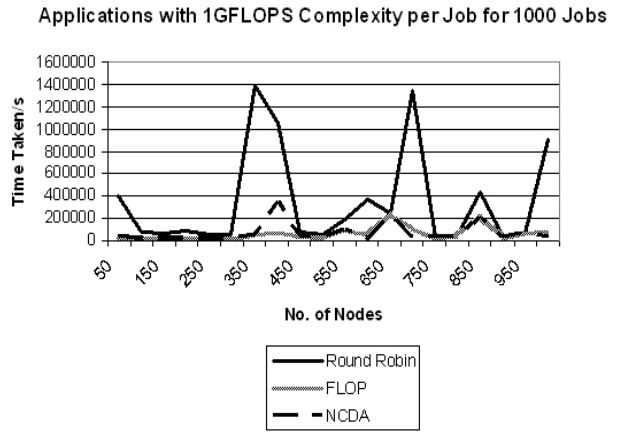

Figure 4: Performance of algorithms for compute intensive applications

Analyzing the results we can see that NCDA provides good performance in environments where network connectivity matters, such as in network intensive (Figure 3) and hybrid applications (Figure 5). It however under-performs for computational complex tasks (Figure 4), in which case the FLOPS based algorithms have the best performance. In the results shown above, the overhead in scheduling is not shown. NCDA has an O(n) scheduling complexity, which means that as the number of the nodes increases, the time to perform a scheduling decision increases proportionally. Additionally there are some overheads involved in determining the bandwidth between the nodes. To handle this we have designed our discovery topology around a two-tier super peer based architecture where sub-grids are formed on the basis of small round trip time (RTT) to limit the effects of network latencies and scale (see section 4.4.).

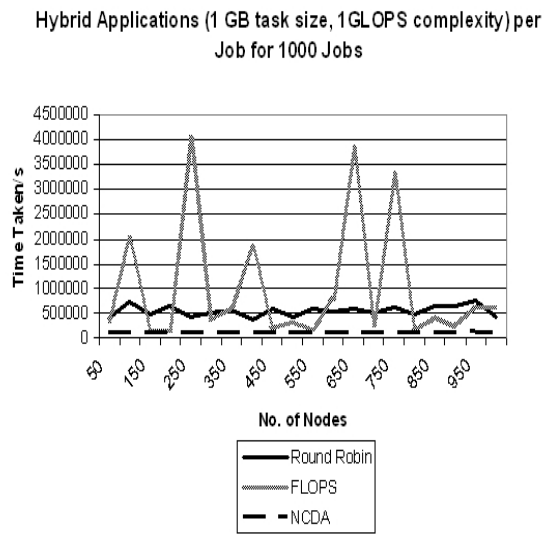

Figure 5: Performance of algorithms for both network and compute intensive applications

**4.4 Resource discovery Architecture**

Our discovery scheme for PhantomOS is an enhancement over the approach of Mastroianni et al. [23]. The enhancements target certain limitations, primarily dealing with the adaptability of the algorithm for hybrid grids and in limiting the overhead of communication between the nodes in a single instance of resource discovery and usage. Certain enhancements deal with limiting the potential for all–to-all communication which plague existing P2P networks.

We introduce a two-tier based super peer architecture. The lowest tier is a machine level granularity sub-grid, which consists of machines that have good network connectivity between them, analogous to a traditional cluster. Each sub-grid is represented by a super-peer, which is the most available machine within the vicinity of the sub-grid. At the top-most tier the granularity is in terms of sub-grids, and these are grouped into regions depending on geographical proximity of the super peers. The regions are represented by a region peer, as shown in Figure 6. A virtual organization (VO) in this system can be at any level: it can consist of individual machines or be an aggregation of entire sub-grids or of entire regions. Interactive applications will be handled at a machine-level VO, whereas large-scale grid applications will require aggregations of entire sub-grids.

The whole concept of a two-tier super peer based system was developed for three main reasons:

- To improve the network usage, by allowing a resource request to propagate to peers in close proximity, thus limiting the overall network traffic, and improving response latency.
- To improve the quality of results, by propagating the request until a suitable resource has been found, while limiting the network traffic as much as possible and
- To provide a scalable and efficient framework for PhantomOS. By dynamically grouping nodes into sub-grids, and clustering sub-grids into regions, QoS is ensured for individual nodes, and the overall network efficiency is enhanced by limiting the flow of resource requests;

The resource discovery mechanism is explained separately in terms of tiers in the following sub-sections of this paper.

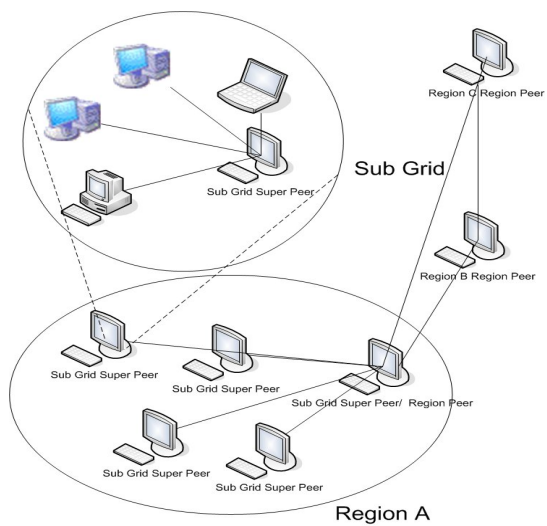

Figure 6: Two Tier Super Peer architecture

**4.4.1 Resource discovery at the Intra-Sub Grid Level**

The sub-grid in our proposed scheme is analogous to a cluster of computers, and is the lowest tier in the system. Resource discovery and brokering is carried out internally in the sub-grid in a semi-centralized fashion. The central server in the sub-grid is the super peer, corresponding to the most available machine in the cluster, and has the responsibility of managing requests and providing a registration interface to new nodes. Upon joining a sub-grid members register their presence with the super peer. When a node of a sub-grid needs a resource, it sends a request query to its super-peer which returns the list of resources matching the user's query constraints. This only happens where matching resources are available. If the super-peer cannot satisfy the request, it then forwards the query request to the region peer. Once the requesting machine has a list of the machines within the sub-grid, the resource broker determines the suitability of the discovered nodes to execute the user application, leading to the eventual submission of the job.

### 4.4.2 Resource Discovery at Intra and Inter-Region Level

If a resource request cannot be satisfied from within the sub grid, the region peer comes into play. The region peer has a notion of the cumulative power of a sub-grid in terms of both data storage for data intensive applications and computational power for compute intensive applications, and based on this it takes a decision on which sub-grids have the required resources to compute the job. The cumulative power of a sub grid is determined by aggregating individual resource descriptions and calculating a theoretical peak. When such sub-grids are found, the job request is forwarded to them and then the resource brokering and scheduling process takes place within the new sub-grid. If the region cannot satisfy the resource requirements it then contacts other regions in a P2P manner.

### 4.4.3 Self-healing Sub-grids and Regions

A self-healing behaviour is crucial in widely distributed architectures such as a Grid environment. This prevents portions of the Grid being crippled when failures occur at crucial nodes, such as hardware or software failures at the super-peer.

To make sub-grids self-healing, a distributed leader election algorithm [24] is deployed to elect a new super peer in a sub-grid. To facilitate the re-generation of data which was stored by the failed super peer, the data is mirrored via erasure coding onto the peers in the sub-grid while a super peer is operational. A new super peer can then regenerate the data from the encoded data. One advantage of modern erasure coding algorithms is that the original data can be regenerated from a portion of the encoded data [25], hence making it suitable for use in widely distributed system. A problem arising from this is the fact that a new node joining the sub-grid will not know the latest super-peer. This is solved using multicast-based discovery within that particular sub-grid.

A region can be made self-healing in the same manner as a sub-grid. A region peer can be selected from amongst the remaining super-peers in case of failure. The same problem arises here as in the sub-grid: how does a new site know which is the latest region peer? We plan to perform discovery at the region-level via the use of location-independent URIs. The current region-peer will always have a specific URI. URI-lookup will be performed via the existing DNS system. Thus new sites can locate the current region-peer and register with it.

## 5. Conclusions and Future Work

PhantomOS is not only aimed at adapting Grid computing for new fields but also at improving facilities for existing Grid users. The goal is to relieve them from many of the problems which are a common way of life within existing infrastructures. Such problems are frequently related to the set up and administration of Grids. The creation of Grid applications and the lack of general fault tolerance within the Grid infrastructure are also issues of concern. PhantomOS is a step towards a "Plug and Play" pervasive grid computing environment. It is designed to support all types of modern computations, including batch and interactive and support the creation of Grids of any architecture. It is based on a self-organizing and self-healing foundation created using the super peer paradigm. The main contribution of this paper is that it presents a novel architecture for the development of Grid OS whilst at the same time commenting on the project's implementation status

Future work includes extending the discovery service to enable self-healing and self organizing behaviour. We also plan to develop a KVM based virtualization engine for PhantomOS to provide security and fine grained resource management to resource owners and privacy to resource users. A generic framework for extending traditional cluster middleware like Condor to add self organizing and fault tolerant behaviour is also required. It is important that PhantomOS includes support for existing grid applications and the associated software such as workflow planners etc. Furthermore we propose that the system should embed the capability for interoperability with existing and emerging Grid infrastructures by making the system compliant to evolving standards in Grid computing. This would include for example, the Web Services Resource Framework (WSRF), as approved by the Open Grid Forum and this will be a major focus of our continuing work.